\documentclass[a4paper,11pt]{article}

\pdfoutput=1

\usepackage{jinstpub} 

\usepackage{lineno}
\usepackage{breqn}
\usepackage{placeins}

\title{\boldmath Setup of Compton polarimeters for measuring entangled annihilation photons}
\def \pathpics {pictures}

\author[a]{D. Abdurashitov,\note{Corresponding author.}}
\author[a,b]{A. Baranov,}
\author[a]{D. Borisenko,}
\author[a]{F. Guber,}
\author[a,1]{A. Ivashkin,}
\author[a]{S. Morozov,}
\author[a,c]{S. Musin,}
\author[a,c]{A. Strizhak,}
\author[a]{I. Tkachev,}
\author[a,c]{V. Volkov,}
\author[a]{B. Zhuikov}

\affiliation[a]{Institute for Nuclear Research RAS, prospekt 60-letiya Oktyabrya 7a, Moscow, Russia}
\affiliation[b]{National Research Nuclear University MEPhI, 31 Kashirskoe Shosse, Moscow, Russia}
\affiliation[c]{Moscow Institute of Physics and Technology , Institutsky lane 9, Dolgoprudny, Moscow region, Russia}

\emailAdd{ivashkin@inr.ru}

\abstract{
An experimental setup  for studying the Compton scattering of annihilation photons in various (entangled and decoherent) quantum states is presented.  Two entangled $\gamma$-quanta with an energy of 511 keV and mutually orthogonal polarizations are produced by positron-electron annihilation in a thin aluminum plate and are emitted in opposite directions. To measure both photons, the setup provides two equivalent arms of  Compton polarimeters. A Compton polarimeter consists of a plastic scintillation scatterer and an array of NaI(Tl) detectors for measuring photons deflected at an angle of 90$^\circ$. The intermediate scatterer of the GAGG scintillator with SiPM readout is inserted into one of the arms to create a tagged decoherence process prior to the measurement of annihilation photons in polarimeters. The performance of Compton scatterers and NaI(Tl) counters is discussed. The polarization modulation factor and the analyzing power of Compton polarimeters are evaluated from the angular distributions of scattered gammas. The Compton scattering of photons in entangled and decoherent states is compared reliably for the first time. 
}


\begin{document}
\maketitle
\flushbottom

\section{Introduction}
\label{sec:intro}
A pair of annihilation $\gamma$-photons with an energy of 511 keV became the first system in which the entanglement of quantum states was experimentally studied \cite{bohm}. As follows from the parity and angular momentum conservation, the photons produced in positron-electron annihilation have mutually orthogonal polarizations. According to the theoretical calculations in \cite{pryce} and \cite{snyder}, this orthogonal polarization leads to the angular correlations of the Compton scattered gammas. Namely, the number of scattered photons depends on the relative azimuthal angle  in the scattered pair and peaks at $90^{\circ}$.   A series of the experiments performed in the second half of 20-th century confirmed the theoretically predicted behavior, see, for example, \cite{kasday}, \cite{bruno}, \cite{langhoff}.  The observed angular correlations are in agreement
with the assumption that the annihilation photons are entangled. 

Strong angular correlations of scattered gammas from the initially entangled pair of annihilation photons imply a completely different scattering kinematics for decoherent photons with the lost entanglement. Since the decoherent photons have no quantum entanglement of the polarization states,  in this case, the absence of azimuthal correlations was assumed \cite{bohm}, \cite{caradonna}. The expected difference in the behavior of Compton scattering kinematics for entangled and decoherent annihilation photons has prompted the development of new approaches in Positron Emission Tomography (PET) \cite{toghyani}, \cite{kozuljevich}. The polarization correlations of annihilation gammas could provide the opportunity to remove the scatter and random backgrounds and to improve the PET image quality \cite{watts}.   Note, that the recently constructed Jagiellonian Positron Emission Tomograph (J-PET) \cite{pet1}, \cite{pet2}  has the possibility of the measurement of the discussed angular correlations, since the double Compton scattering of photons in J-PET plastic scintillators  enables to determinate the linear polarization of the primary photons. 

The situation became rather vague a  few years ago after the appearance of a theoretical paper \cite{hiesmayr}, where the application of open quantum formalism revealed the same Compton scattering kinematics for both entangled and decoherent states. Accordingly, previous experimental studies are considered incomplete. A direct experimental comparison of the Compton scattering kinematics for these two quantum states is required. 

First attempt to compare the Compton scattering of entangled and decoherent annihilation photons was recently made using the PET demonstrator \cite{watts}.  In this experiment, photon pairs were subjected to a decoherence process  with a passive scattering medium in the path of one of the original gamma. Unfortunately, large statistical uncertainties and low sensitivity to polarization do not allow to draw an unambiguous conclusion. As pointed out in \cite{watts}, dedicated measurements with increased statistical accuracy are needed to clarify the  relevance of quantum entanglement to the future generation of positron emission tomography. 

Meanwhile, Compton scattering of entangled or decoherent photons is not limited to PET applications and is a fundamental process that is still poorly understood. The dedicated study requires an experimental setup with the high sensitivity to measured polarization states and reliable identification of quantum state. The experimental setup described below meets these requirements. The setup consists of two groups of Compton polarimeters with a high analyzing power and a large solid angle coverage. Prescattering in the GAGG scintillator provides the decoherence process and the capability to tag the decoherent events.
The azimuthal symmetry of the setup eliminates systematic errors and ensures the reliability of the experimental results.

\section{Principles of polarization measurements of annihilation photons}

Compton polarimeters utilize the angular dependence of the scattering probability on the direction of photon polarization. It follows from the differential cross-section for Compton scattering of linearly polarized photons given by the Klein-Nishina formula \cite{nishina}:

\begin{equation}
d\sigma/d\Omega=r_e^2/2\cdot(E_1/E)^2\cdot(E_1/E+E/E_1-2\sin^2{\theta}\cdot\cos^2{\phi}),
\label{eq:kl-ni}
\end{equation}
where ${r_e}$ is the classical electron radius, $E$ is the energy of the incident photon, $m_e$ is an electron mass, $E_1$ is the energy of the scattered photon ${E_1=E\cdot m_e/(m_e+E\cdot(1-\cos{\theta})})$, $\theta$ is the angle between the incident and scattered photons momenta forming the scattering plain, and $\phi$ is the angle between the  scattering plane  and the direction of polarization of the incident photon. As follows from Eq. \ref{eq:kl-ni}, for given scattering angle $\theta$ the cross-section is maximum  for $\phi = 90^{\circ}$.  This feature is widely used to measure the polarization of high energy photons using Compton polarimeters.

\begin{figure}[htbp]
\centering 
\includegraphics[width=.40\textwidth]{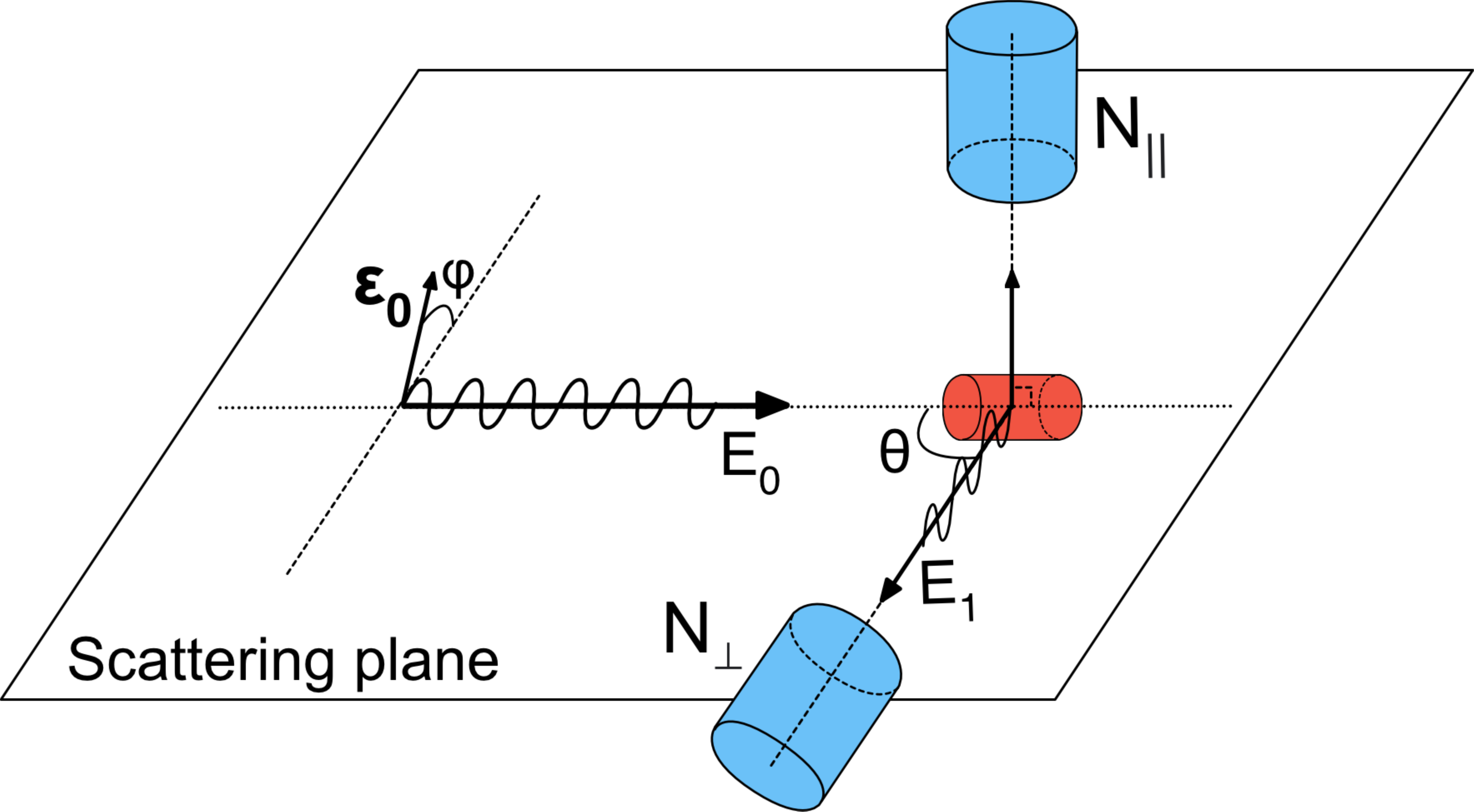}
\qquad
\includegraphics[width=.54\textwidth]{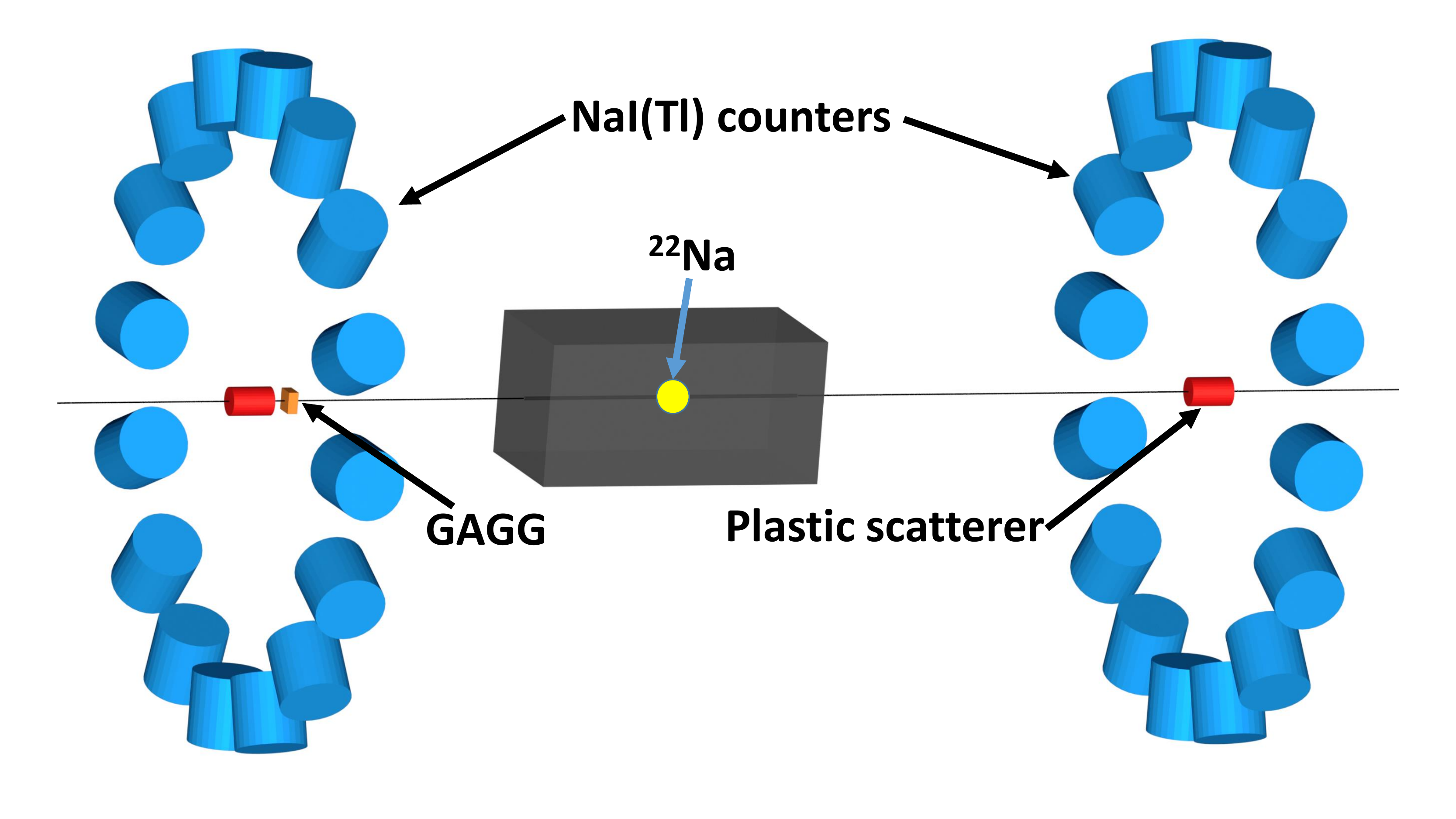}
\caption{ Left – scheme of an elementary Compton polarimeter consisting of a scatterer and two orthogonal detectors of scattered gammas.   $E_0$ and $E_1$ are the energies of initial and scattered gammas, respectively, $\phi$ is the angle between initial polarization vector $\varepsilon_0$ and the scattering plane.
Right – layout of experimental setup, that includes two equivalent arms with 16 elementary Compton polarimeters in each arm. An intermediate scatterer of GAGG scintillator is placed in one arm to produce the decoherent photons. }
\label{fig:1}
\end{figure}

The schematic structure of an elementary Compton polarimeter is shown in figure \ref{fig:1}, left. It consists of a Compton scatterer and two mutually orthogonal detectors of scattered photons. Analyzing power $A$ of a Compton polarimeter is defined as $A(\theta)=\frac{N_\perp-N_\parallel}{N_\perp+N_\parallel}$, where ${N_\perp}$ $({N_\parallel})$ denotes the number of detected events in counters located perpendicular (parallel) to the polarization of initial photons. Using Eq. \ref{eq:kl-ni} the analyzing power is obtained as \cite{knights}:
\begin{equation}
A(\theta)=\frac{\frac{d\sigma}  {d\Omega}(\theta,\phi=90^\circ)-\frac{d\sigma}{d\Omega}(\theta,\phi=0^\circ)}{\frac{d\sigma}  {d\Omega}(\theta,\phi=90^\circ)+\frac{d\sigma}{d\Omega}(\theta,\phi=0^\circ)} 
=\frac {\sin^2{\theta}}
{E_{1}/E+E/E_{1}
-\sin^2{\theta}}
\label{eq:apow}
\end{equation}
As follows from Eq. \ref{eq:apow},  for a given energy of initial gamma, the analyzing power depends on  the scattering angle.

In the case of annihilation photons with the energy equal to the electron mass, Eq. \ref{eq:kl-ni} can be rewritten as: 

\begin{equation}
\frac{d\sigma}{d\Omega}= \left( \frac{d\sigma}{d\Omega} \right) _{NP}\cdot(1-\alpha(\theta)\cos(2\phi))
\label{eq:polsec}
\end{equation}
where 
\begin{equation}
\left( \frac{d\sigma}{d\Omega} \right) _{NP}=r_e^2\cdot\epsilon^2(\epsilon+1/\epsilon-\sin^2{\theta})
\label{eq:npsec}
\end{equation}
is the differential cross-section for the non-polarized photons and ${\epsilon=E_1/E}$. 
Parameter 
$\alpha(\theta)=
\sin^2{\theta}/(\epsilon+1/\epsilon-\sin^2{\theta})$
defines behaviour of the azimuthal angular distributions of scattered gammas with respect to the polarization plane of the initial annihilation photons and characterizes the sensitivity of Compton polarimeters to the measured polarization.
One can see, that for completely polarized photons the analyzing power coincides with the parameter 
 $\alpha(\theta)$ in Eq. \ref{eq:polsec} and reaches maximum of $A=0.69$ at $\theta=82^\circ$ for annihilation photons with the energy of 511 keV.
 
In the case of two entangled annihilation photons with mutually orthogonal polarizations, the probability of Compton scattering at angles $\theta_1, \theta_2$  is given by the following expression \cite{caradonna}, \cite{hiesmayr}:
\begin{equation}
P(E_1,E_2,\Delta\phi)=\left( \frac{d\sigma}{d\Omega_1} \right) _{NP} {\left( \frac{d\sigma}{d\Omega_2} \right) _{NP}\cdot(1-\alpha(\theta_1)\alpha(\theta_2)\cos(2\Delta\phi))}
\label{eq:prob}
\end{equation}
where $ \Delta\phi$ is the angle between scattering planes, parameters $\alpha(\theta_1)$ and $\alpha(\theta_2)$ are related to the first and second photons and have the same meaning as in Eq. \ref{eq:polsec}. Since the entangled gammas have 100$\%$ mutually orthogonal polarizations, $\alpha(\theta_1)$ and $\alpha(\theta_2)$ coincide with the analyzing power of the corresponding Compton polarimeter.

The sensitivity to the polarization measurements for this system of two photons is characterized by the polarization modulation factor $\mu$, which is defined as: 

\begin{equation}
\mu=\frac{P(\Delta\phi=90^\circ)-P(\Delta\phi=0^\circ)}{P(\Delta\phi=90^\circ)+P(\Delta\phi=0^\circ)}
\label{eq:modul}
\end{equation}                                                                                                              
where $P(\Delta\phi = 90^\circ)$ and $P(\Delta\phi = 0^\circ)$ are the probabilities of detecting gammas scattered in orthogonal and parallel directions, respectively. As follows from Eq. \ref{eq:prob} and Eq. \ref{eq:modul}, the modulation factor is the product of analyzing powers of two Compton polarimeters used for the measurements of pair of annihilation photons, $\mu=\alpha(\theta_1)\alpha(\theta_2)$. It reaches a maximum of 0.48 for $\theta_1=\theta_2 =82^\circ$ \cite{pryce}, \cite{snyder} in accordance with the values of analyzing powers of two individual Compton polarimeters. Due to the finite geometry of the detectors, the modulation factor in a real experimental setup is lower.

\section{Experimental setup}
\begin{figure}[htbp]
\centering 
\includegraphics[width=.58\textwidth]{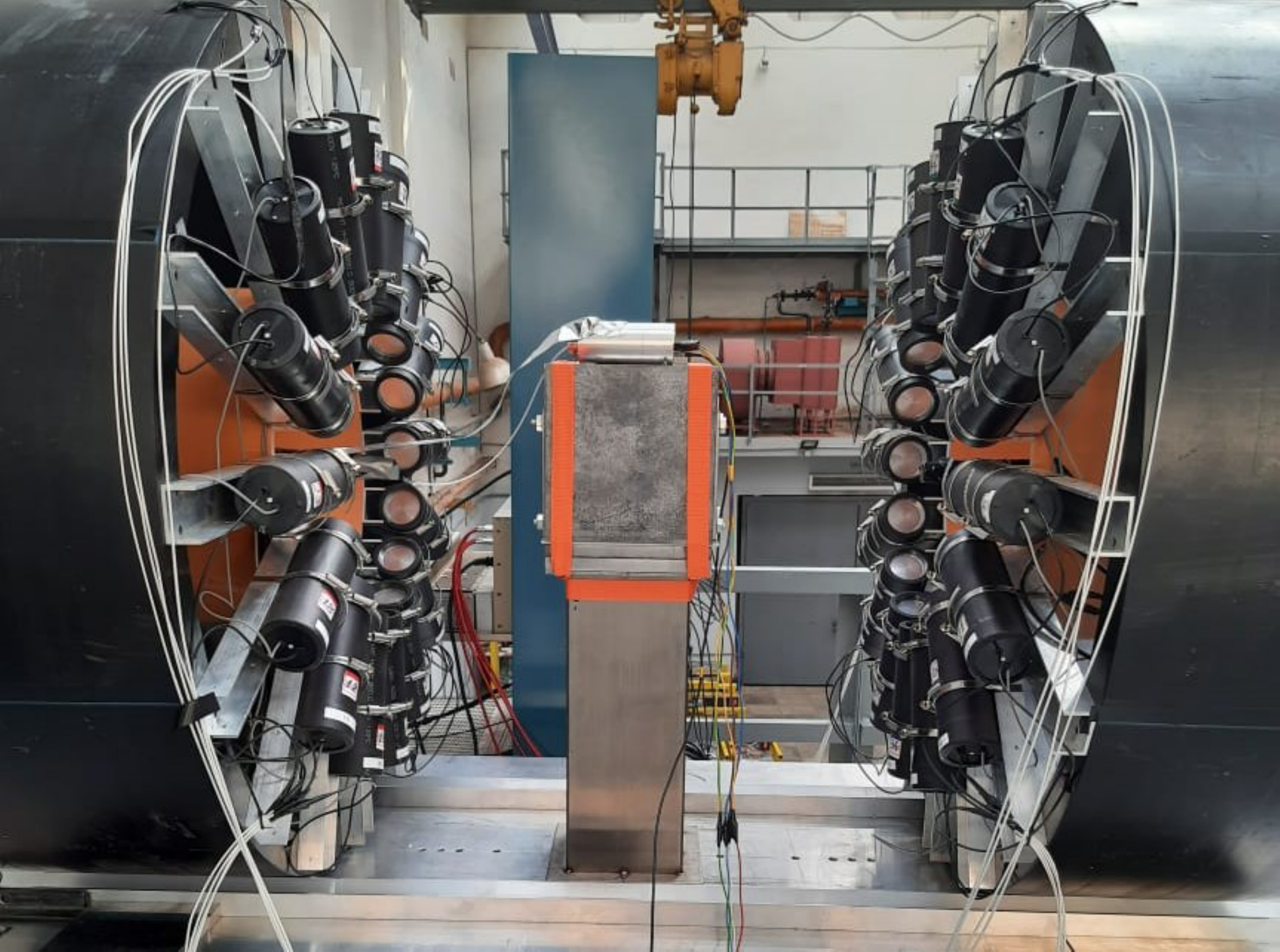}
\qquad
\includegraphics[width=.325\textwidth]{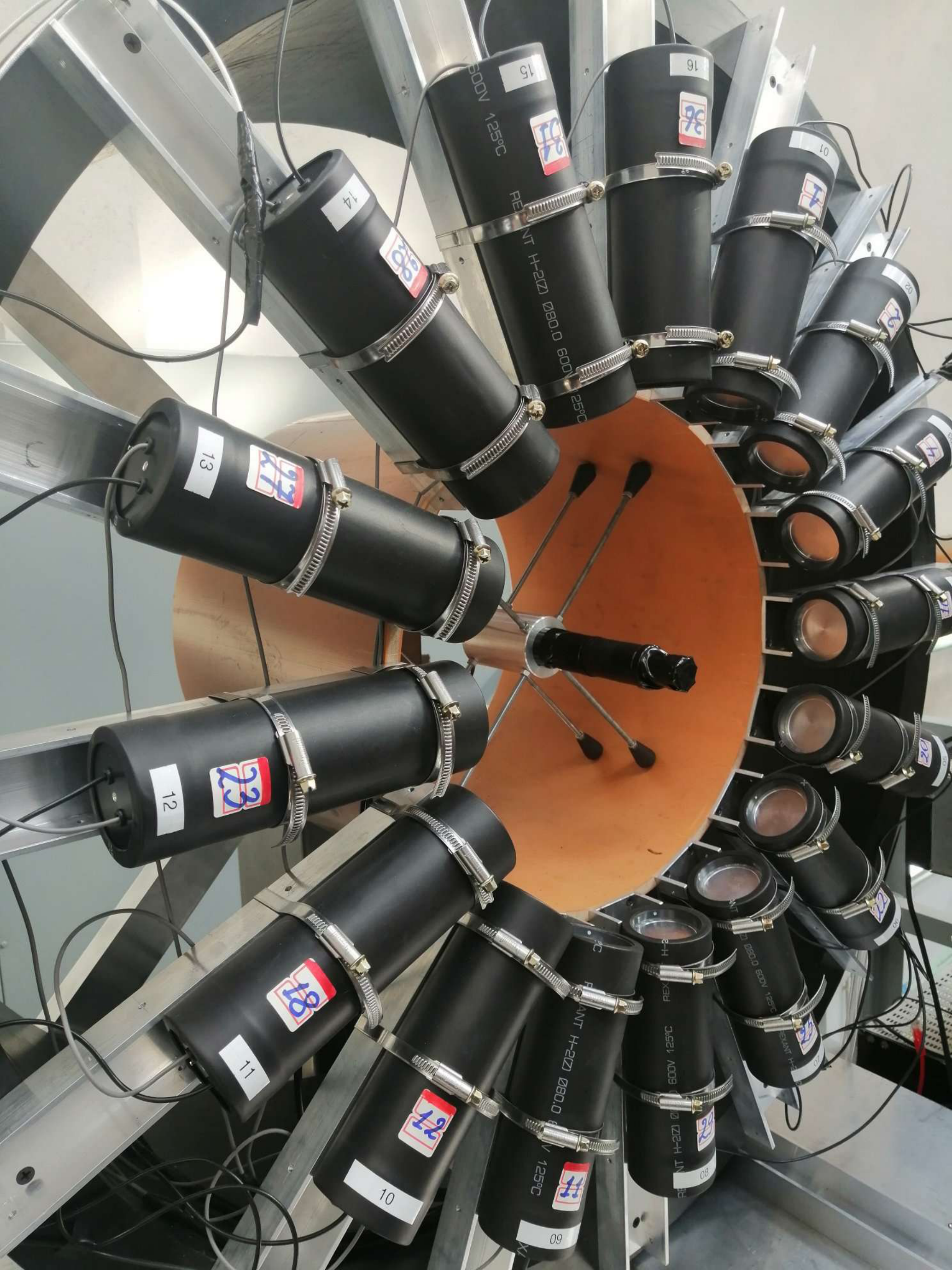}
\caption{ Photo of the experimental setup (left) and photo of one arm of the setup (right). Each arm includes a plastic scatterer on setup axis and 16 NaI(Tl) counters orthogonal to the axis. The $^{22}$Na source of positrons located in lead collimator is placed between the arms. }
\label{fig:2}
\end{figure}

The scheme of the experimental setup for studying Compton scattering of entangled and decoherent annihilation photons is presented in figure \ref{fig:1}, right. It consists of two arms of Compton polarimeters and a source of annihilation gammas between them. Each arm includes the scatterer of plastic scintillator and a set of 16 NaI(Tl) detectors of scattered gammas, symmetrically placed around the scatterer and perpendicular to the setup axis. Each pair of mutually orthogonal NaI(Tl) counters and the corresponding scatterer compose an elementary Compton polarimeter. As a result, the setup comprises 32 polarimeters, 16 on each side. The detectors of scattered gammas are \o50{$\times$}50 mm$^2$ NaI(Tl) cylinders with Hamamatsu R6231 PMT readout. They are placed at the distance of about 200 mm from the setup axis. Each scatterer is a \o20{$\times$}30 mm$^2$ cylinder of plastic scintillator with Hamamatsu R7525 PMT readout. The distance between the arms of the setup is about 70 cm. The photos of the full setup and of the one of the arms are shown in figure \ref{fig:2}.

To produce decoherent pairs of annihilation photons, an intermediate scatterer of 7 mm thick GAGG scintillator is placed near one of plastic scatterer. High light yield of GAGG allows to detect the energy depositions of recoil electrons with a threshold of a few keV. Registration of gamma interaction in intermediate scatterer indicates that the initially entangled pair of photons has become decoherent. To minimize the interaction rate in passive material of intermediate scatterer, the light  of GAGG scintillator is read by silicon photomultiplier (SiPM) matrix with  $\sim$1 mm thickness. The SiPM Hamamatsu MPPC S14161-3050HS-04 with transverse sizes of 13{$\times$}13 {mm$^2$} was selected due to its high gain of about $2.5\cdot10^6$ and quantum efficiency of about 50\%. 

Radioactive $^{22}$Na source of positrons with activity $\sim$50 MBq was prepared by irradiating a 1 mm thick aluminum plate at INR RAS isotope production facility \cite{source} with 130-MeV protons. The plate is fixed transversely inside a hollow aluminum cylinder with an inner diameter of 5 mm. The source is placed in the center of cubic 20{$\times$}20{$\times$}20 cm$^3$ lead collimator with a 5 mm diameter horizontal hole. Two 511 keV $\gamma$-photons with opposite momenta are produced in positron-electron annihilation in aluminum plate. 
In principle, an alternative annihilation into 3$\gamma$ is also possible but is strongly suppressed for the following reasons. Since the annihilation occurs in metal, the positrons are immediately thermalized \cite{garwin} and can form intermediate positron-electron bound state of positronium in singlet (parapositronium) or triplet (orthopositronium) states. Exchange collisions with electrons on the Fermi surface (pick-off process) convert any triplet state to singlet. In this pick-off process the positron of orthopositronium suffers 2$\gamma$-annihilation, as in the case of parapositronium. On the other hand, according to \cite{positron}, positrons in metals are effectively screened by conduction electrons and the formation of positronium is effectively suppressed. As a result, positron-electron annihilation mostly proceeds directly via $e^+e^- \rightarrow 2\gamma$ process and the background 3$\gamma$ annihilation in aluminum is negligible. 

The  $^{22}$Na source is located on the setup axis 10 cm closer to the arm with the GAGG scintillator to ensure that the first interaction occurs in the intermediate scatterer. If no interaction in GAGG is detected a pair of photons is considered to be entangled. 

 Rotational symmetry of the setup provides a few essential advantages. First, it allows to measure polarizations in opposite azimuthal directions without changing the setup geometry. Second, 16 Compton polarimeters in each arm provide large solid angle coverage. It is especially important for collection of data with low count rate, which is the case of decoherent events with double Compton scattering in GAGG and plastic scintillators. Third, due to the azimuthal symmetry of the setup each NaI(Tl) counter measures vertical or horizontal components of polarization depending on the orientation of scattering plane. As a result, possible systematic errors caused by different efficiencies and inaccuracies in locations of NaI(Tl) counters are compensated by the symmetry of the setup design. 
 
The data acquisition system is based on ADC64s2 board developed by AFI Electronics \cite{afi}, which includes 64-channels 12-bit 62.5 MS/s ADCs. The operational amplifiers with the shaping time of about 100 ns are incorporated in each input of ADC board for two reasons. First, the ADC sampling time of 16 ns is too long for the fast signals of plastic scintillators. Therefore, the signals must be appropriately shaped. Second, photomultipliers Hamamatsu R6231 used for readout of NaI(Tl) counters have rather low gain of 2.7x10$^5$ and require an additional amplification for low energy signals.  The waveforms digitized by the ADC are used to evaluate both amplitude and timing information.

The event trigger is organized by a coincidence of signals in two plastic scatterers. This means that only a minor fraction of recorded events has signals in NaI(Tl) counters, while most events represent the gamma scattering in 4$\pi$ solid angle. Such loose trigger conditions allow cross-calibration of on-axis detectors as well as data quality studies.

\section{Detector performances}

\subsection{ADC waveform analysis}
\begin{figure}[htbp]
\centering 
\includegraphics[width=.47\textwidth]{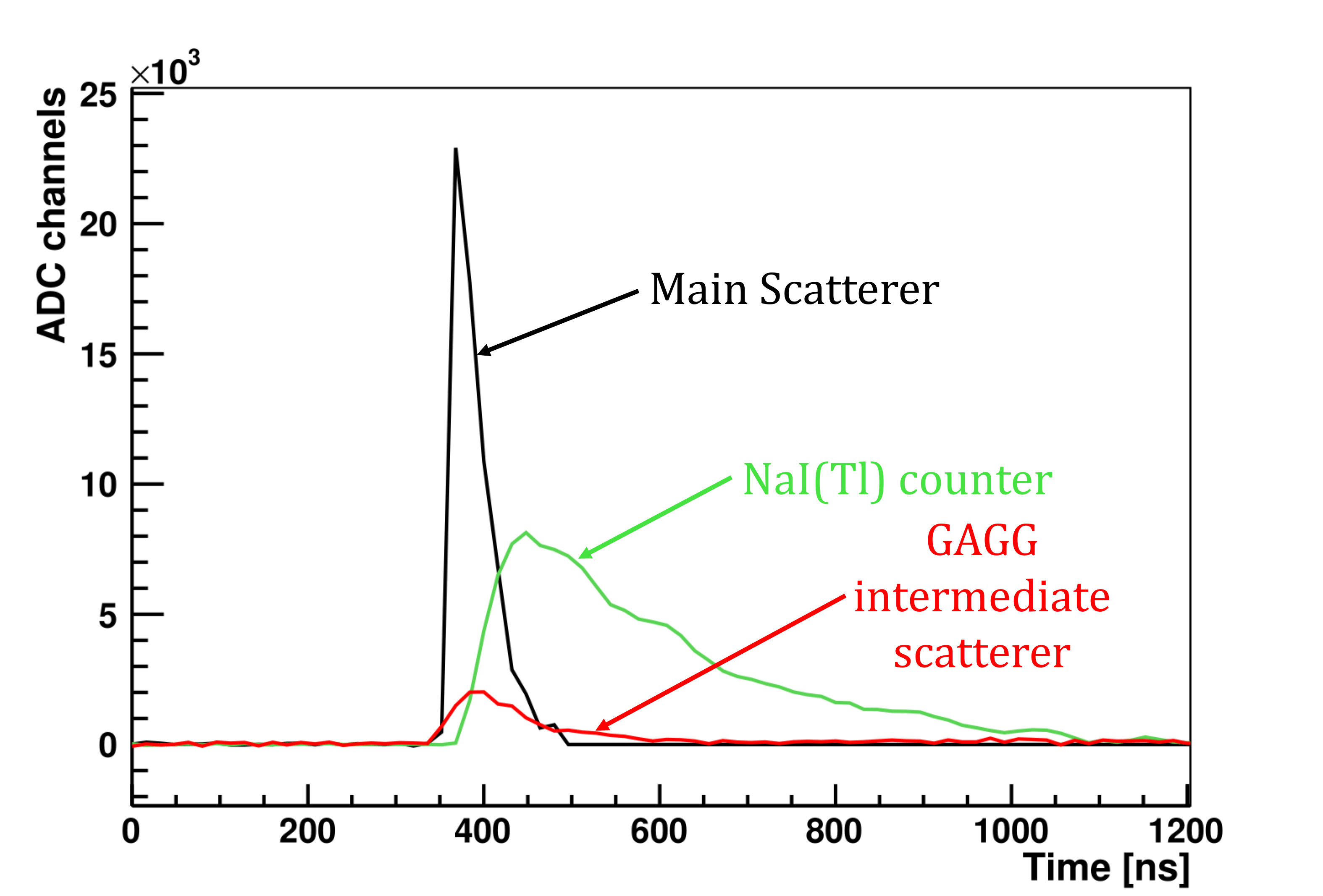}
\qquad
\includegraphics[width=.47\textwidth]{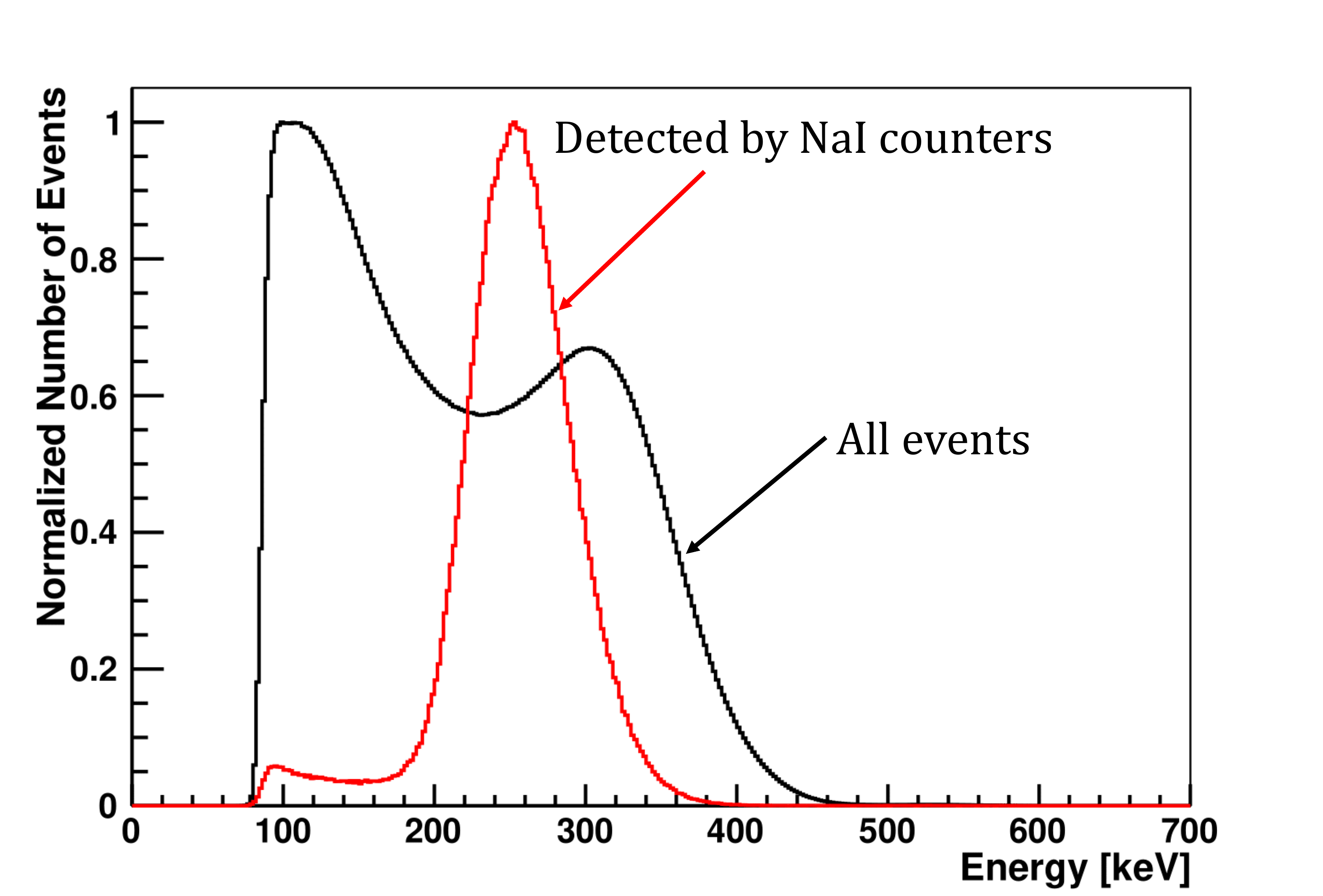}
\caption{ Left - typical signal waveforms in plastic scatterer (black line), in GAGG intermediate scatterer (red line) and in NaI(Tl) counter (green line). Right - normalized energy spectra in plastic scatterer for all events (black line) and detected by NaI(Tl) counters (red line).} 
\label{fig:3}
\end{figure}

Each detector in the setup provides the amplitude and time spectra evaluated from the ADC signal waveforms. The typical examples of waveforms are presented in figure \ref{fig:3},  left. Plastic scintillation scatterers have the shortest signal lengths of about 100 ns, that reflects the shaping time of amplifiers. The signal length for scattered photon detectors exceeds 600 ns and corresponds to the decay time of NaI(Tl) scintillator. The amplitudes are obtained by integrating the waveforms  over the time gate corresponding to the duration of the signals. 

Figure \ref{fig:3}, right shows the amplitude spectra in plastic scatterer for two groups of events. The black line corresponds to the energy spectrum of recoil electrons for Compton scattering  into the full solid angle. This spectrum resembles a typical Compton scattering of annihilation photons in plastic scintillators, see, for example \cite{JPET}.  The red line corresponds to the energy spectrum of the subset of events when  the scattered photons were  registered by NaI(Tl) counters. In Compton scattering of annihilation photons at 90$^\circ$ the energy of recoil electrons equals to the energy of scattered photon.  Therefore, the peak in the energy spectrum corresponds to 255.5 keV and is used for calibration of plastic scatterers every 2 hours of data acquisition.

\begin{figure}[htbp]
\centering 
\includegraphics[width=.47\textwidth]{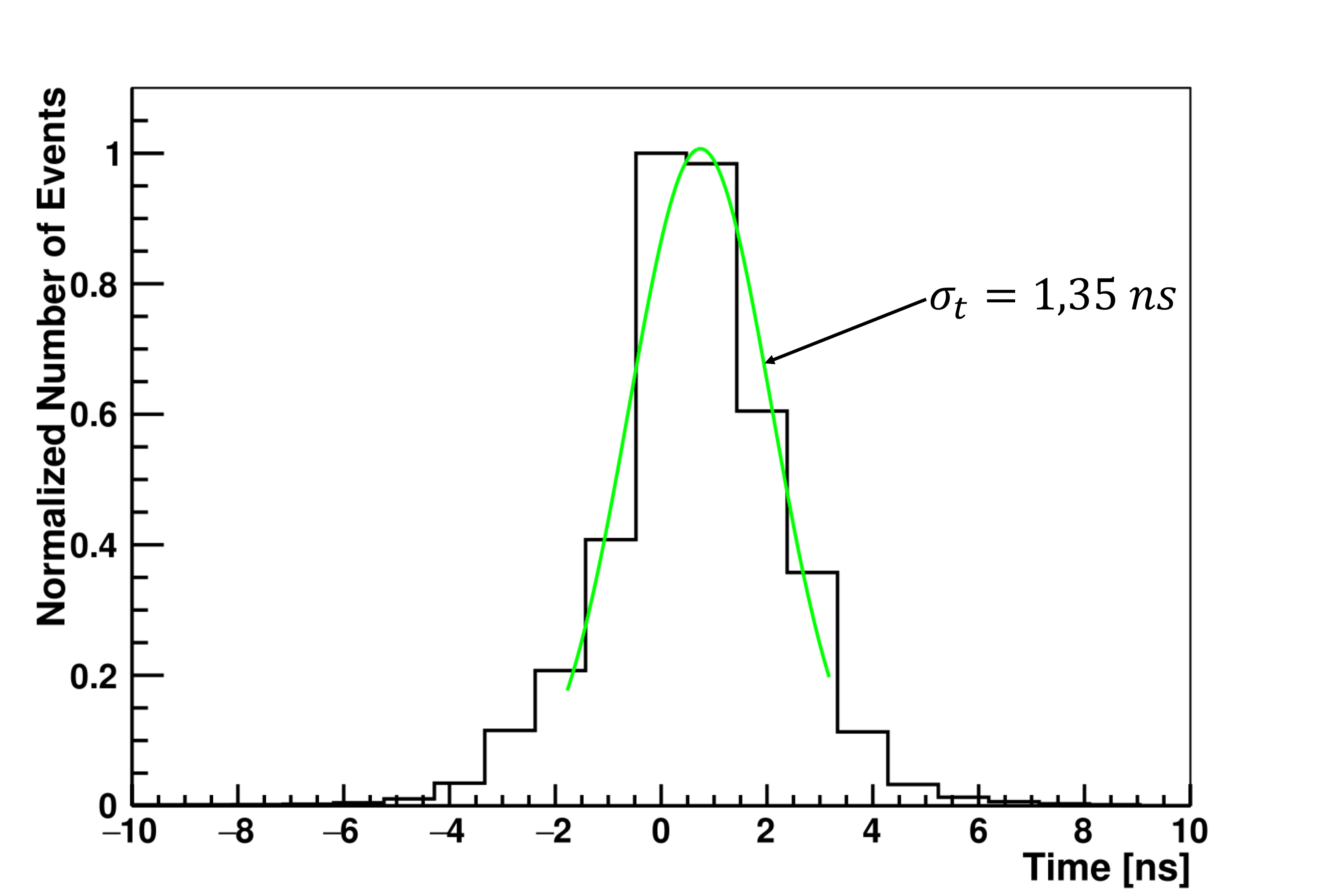}
\qquad
\includegraphics[width=.47\textwidth]{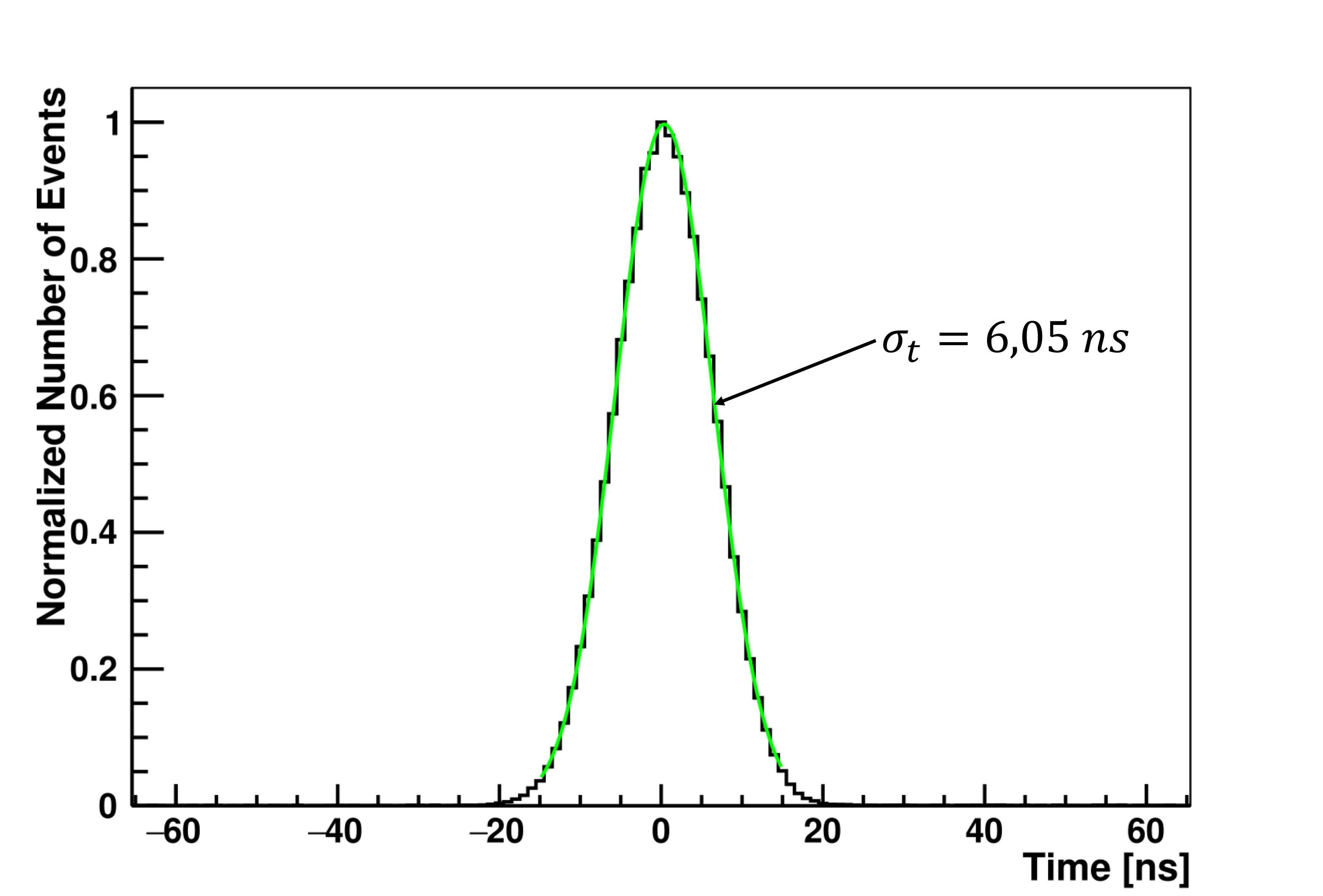}
\caption{ Left - time coincidence spectrum between two plastic scatterers. Right - time coincidence spectrum between NaI(Tl) counter and the plastic scatterer of the same arm.}
\label{fig:4}
\end{figure}

The same signal waveforms were used for the extraction of time information in different detectors. Signal time was calculated as the average time of waveform points with amplitudes, larger than 10 percent of the maximum amplitude: $t_{signal} = \frac{\sum a_i\cdot t_i}{\sum a_i}$, where $t_i$ and $a_i$ are time and amplitude of a corresponding waveform point. Figure \ref{fig:4}, left shows the time coincidence spectrum for the signals from plastic scatterers. The width of this spectrum $\sigma_t$=1.35 ns yields the time resolution of single plastic scatterer $\sigma_{1t}=\sigma_t/\sqrt{2}$=0.95 ns,  which is quite impressive considering that the ADC's 16 ns sampling time is an order of magnitude longer.


Time coincidence spectrum for the signals in NaI(Tl) counters and plastic scatterers is presented in figure \ref{fig:4}, right. Time resolution of counters is about $\sigma_t\approx$ 6 ns and reflects long $\sim$ 250 ns decay time of NaI(Tl) scintillator. 
These time spectra are used for suppression of accidental background that is non-negligible in the long ADC time gate of about $\sim0.6 \mu$s for NaI(Tl) signals.

\subsection{NaI(Tl) counters}

\begin{figure}[htbp]
\centering 
\includegraphics[width=.47\textwidth]{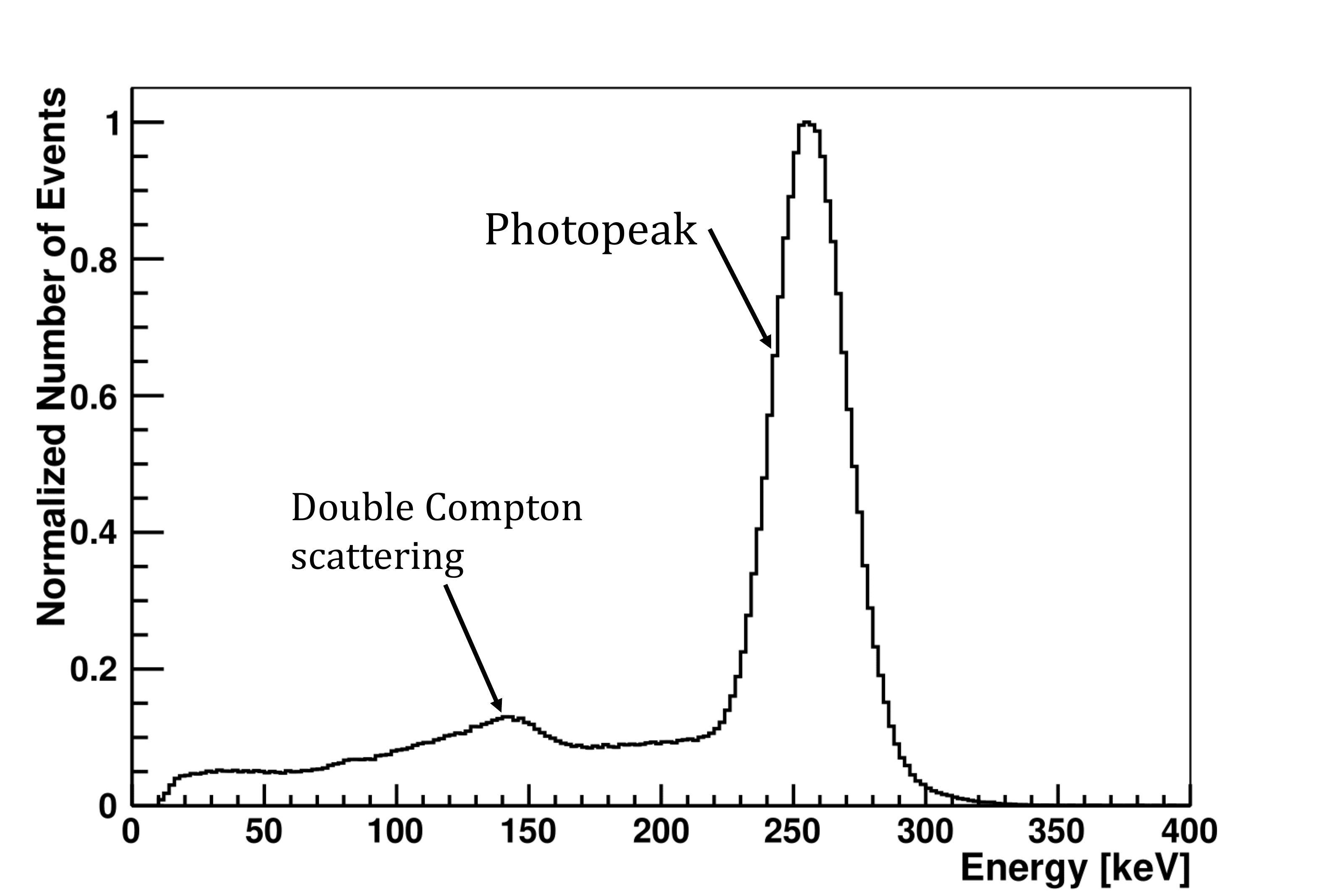}
\qquad
\includegraphics[width=.47\textwidth]{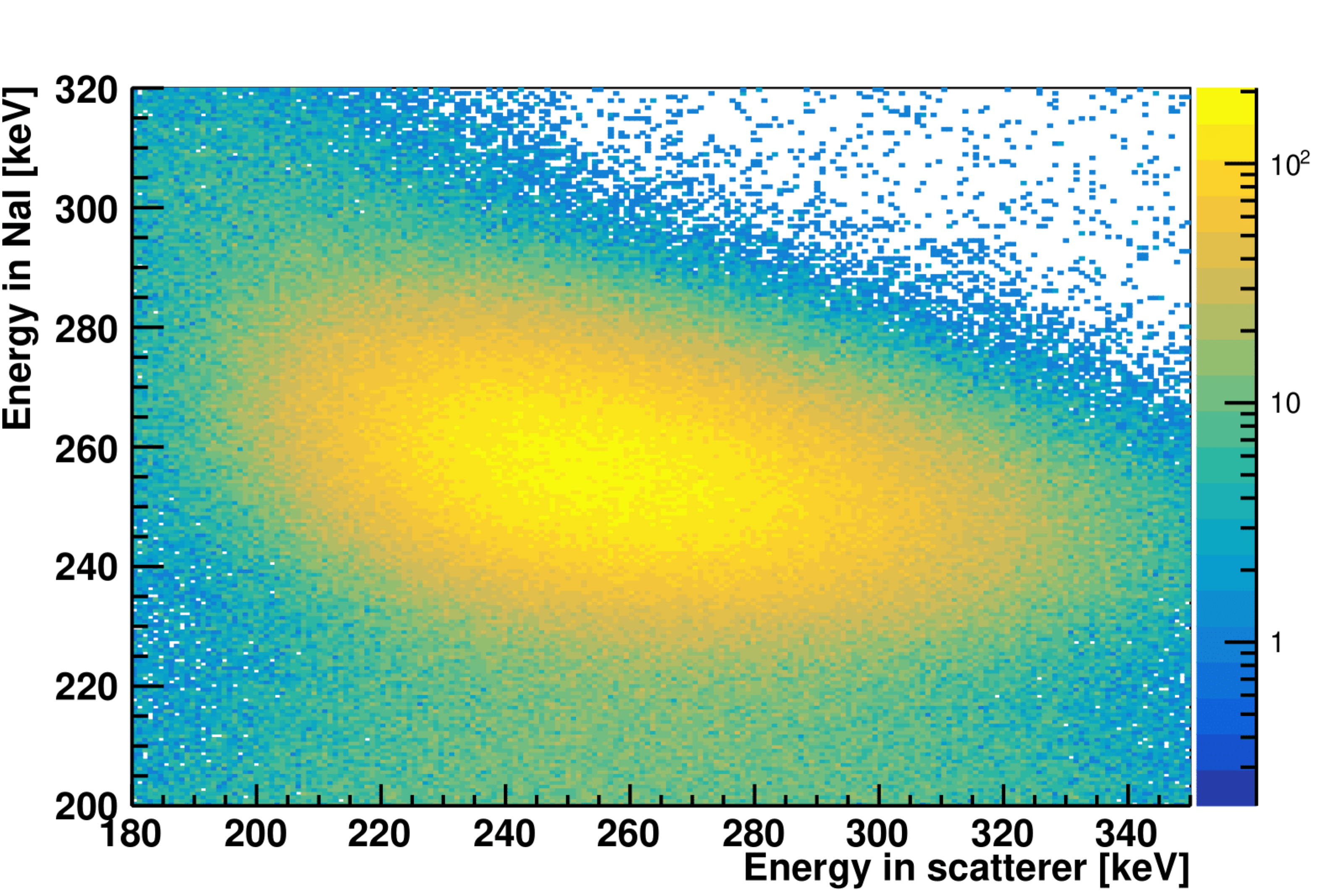}
\caption{ Left - the energy spectrum in NaI(Tl) counter. Right - correlation between energies in plastic scatterer and in NaI(Tl) counters.}
\label{fig:5}
\end{figure}

Compton scattering angles are mainly determined by the azimuthal arrangement of NaI(Tl) counters around the setup axis.  At the same time, the scattering kinematics can be distorted by several parasitic processes, such as double scattering in on-axis detectors or the contribution of 1275 keV nuclear gamma rays emitted by the $^{22}$Na source simultaneously with annihilation photons. These processes are easily distinguishable in the NaI(Tl) energy deposition spectra. Figure \ref{fig:5} presents the energy spectrum in NaI(Tl) counters.
 The left long tail of the spectrum corresponds mainly to the Compton scattering in NaI(Tl) counters.  Here, a small peak is due to double Compton scattering events in on-axis detectors.
The prominent peak in NaI(Tl) spectrum corresponds to the full energy absorption of scattered gamma. This photopeak is centered at an energy close to half the energy of initial annihilation photon, which corresponds to the Compton scattering at $90^\circ$. For physical analysis, only events in the photopeak are used.  This peak is also used for energy calibration of the NaI(Tl) counters every 2 hours of data acquisition.

It should be noted that the contribution of NaI(Tl) counter`s energy resolution to the width of the photopeak is much smaller than the energy spread of registered scattered gamma. According to the Monte Carlo simulation of the setup with realistic geometry, the NaI(Tl) counters detect photons with scattering angles in the range of $80^\circ-100^\circ$ that corresponds to energy deposition from 235 keV to 280 keV.  This is illustrated by the experimental correlation between the energy of recoil electrons in a plastic scatterer and the energy deposition in the NaI(Tl) counters of the same arm, see figure
\ref{fig:5}, right. The correlation confirms the wide range of  photon scattering angles recorded in NaI(Tl) counters.

\subsection{Intermediate GAGG scatterer}

 The intermediate GAGG scatterer is the key element of the setup that separates events into tagged  decoherent or entangled quantum states.  It is located next to one of the plastic scatterers and is the closest detector to the $^{22}$Na source. An interaction in the GAGG scintillator means that a pair of initially entangled annihilation photons  has undergone a decoherence process. Therefore, the reliability of interaction identification in the intermediate scatterer is the most important feature of the setup. Time and amplitude of the  signal are used to identify the interaction in the GAGG scintillator.

\begin{figure}[htbp]
\centering 
\includegraphics[width=.47\textwidth]{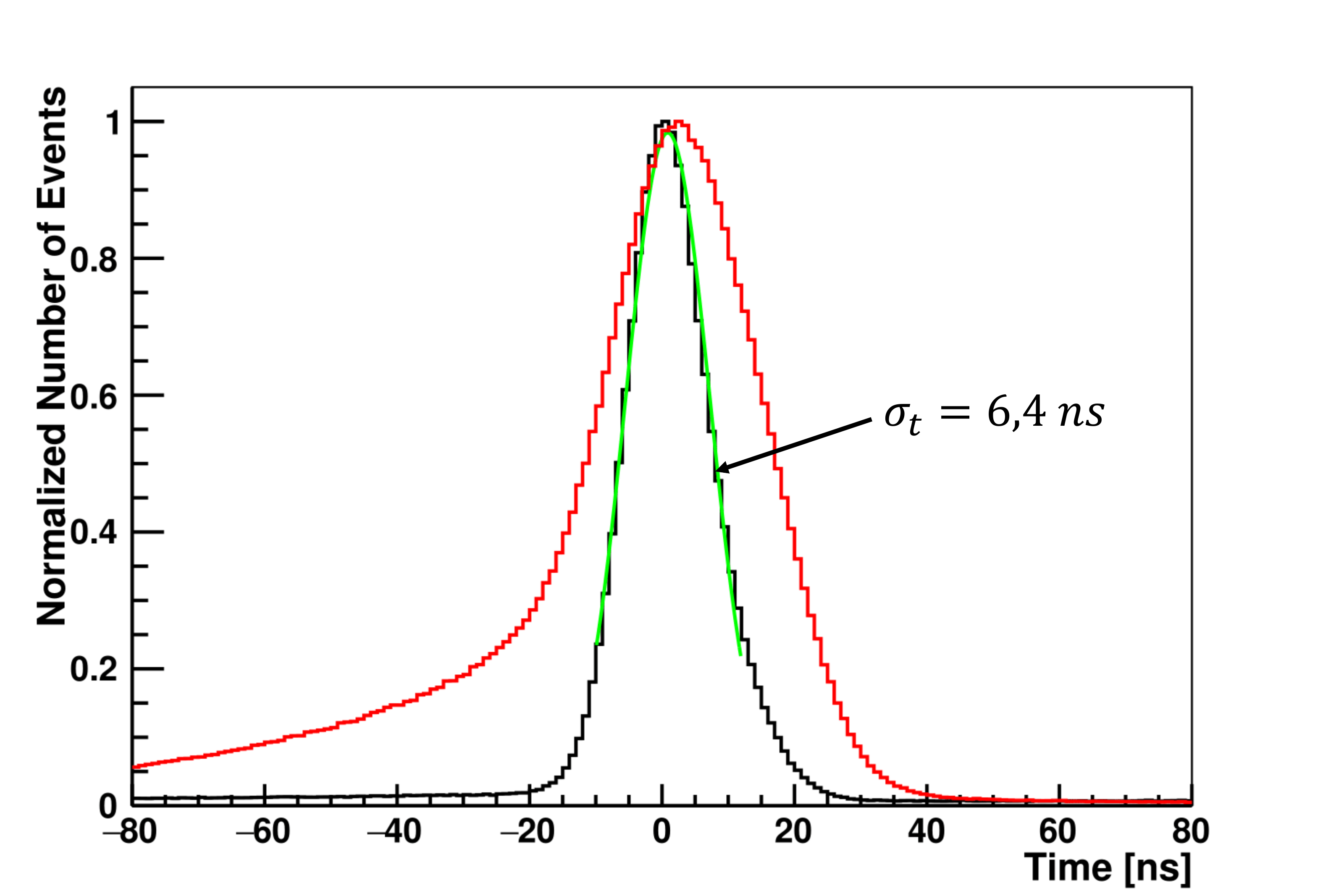}
\qquad
\includegraphics[width=.47\textwidth]{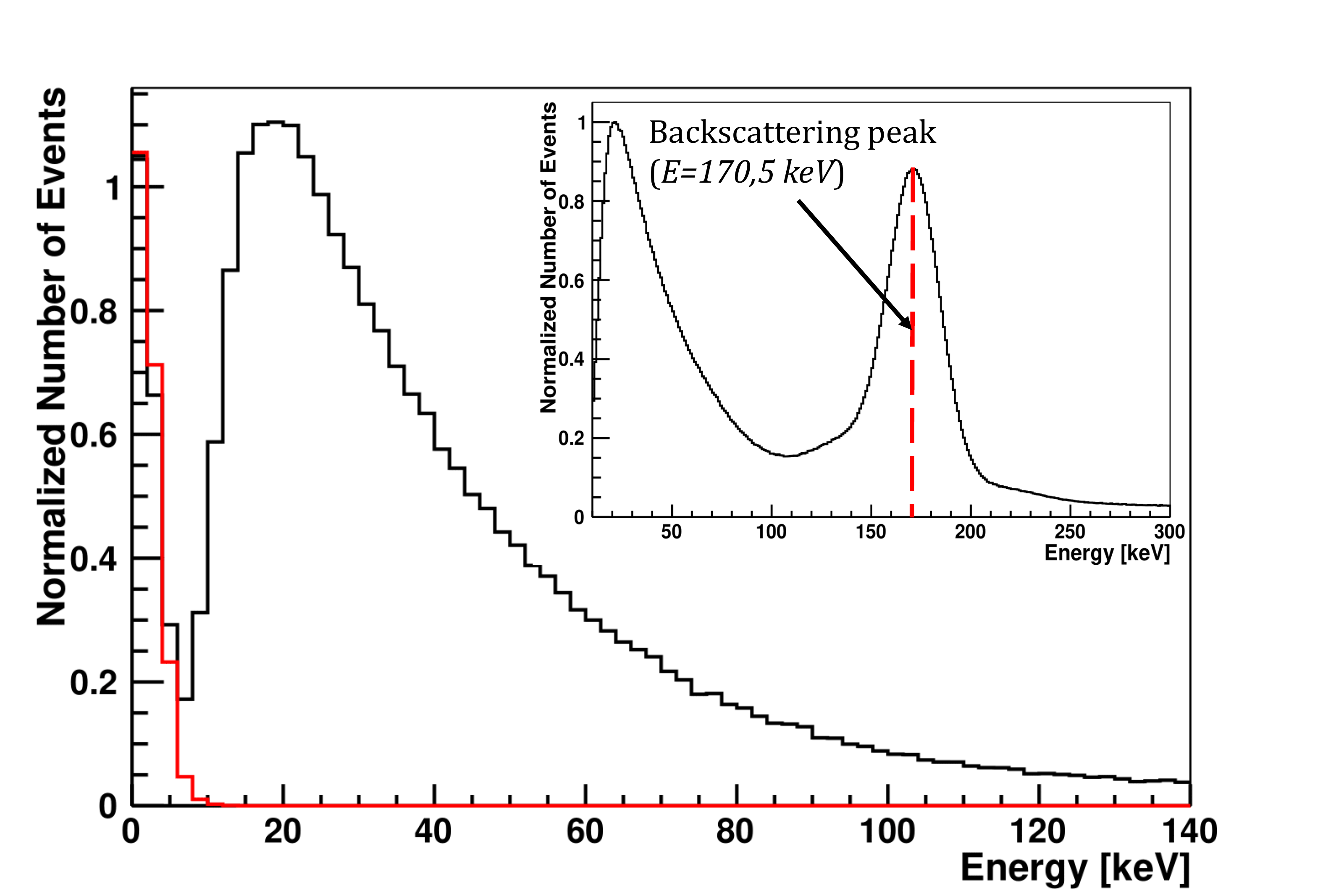}
\caption{ Left - time coincidence spectra for signals in GAGG and plastic scatterers.  GAGG energy deposition in ranges 10-50 keV (red line) and 50-140 keV (black line) is selected. Right - the energy spectra in GAGG scatterer for events outside the true time coincidence window (red line) and  inside the true time coincidence peak (black line). The events with hits in NaI(Tl) counters are selected here. Insert shows the extended GAGG energy spectrum  for all events,  regardless of the hits in NaI(Tl) counters. The prominent peak from photons backscattered by the adjacent plastic scatterer  is used for energy calibration of GAGG.}
\label{fig:6}
\end{figure}

The time coincidence spectra for signals from intermediate and plastic scatterers are presented in figure \ref{fig:6}, left, for two cases.   The red line shows the time spectrum for the lowest, $[10, 50] $ keV, energy deposition range  in the GAGG. In this case the scattering angles are small and the influence on the momentum of initial annihilation photon is minimal. A fairly wide and asymmetric peak in the time spectrum reflects a significant influence of electronic noise on time resolution, as well as the dependence of  time on the signal amplitude. An event is considered decoherent if the coincidence time is in $[-40, 40]$ ns window. 

The black line in figure \ref{fig:6}, left panel, represents the time coincidence spectrum for events with energy deposition in intermediate scatterer in the range of $[50, 140]$ keV. As can be seen, for higher energies the time peak is symmetrical with the width of about $\sigma_t=6.4$ ns. 

Figure \ref{fig:6}, right panel, shows the energy spectra in GAGG scintillator for events in true time coincidence window (black line) and outside this window (red line).  Here, events are selected that hit the NaI(Tl) counters. It can be seen that the equivalent electron noise energy (red line) is less than 10 keV. Therefore, a system of two photons is considered decoherent if energy deposition in GAGG exceeds 10 keV and the coincidence time is in $[-40, 40]$ ns window. If the identified energy deposition in GAGG is below 10 keV and events  are outside the true time coincidence window, the photons are regarded as an entangled pair.  The obtained time and energy spectra in GAGG detector confirm that the setup reliably distinguishes decoherent pairs of photons. 

 The energy calibration of the GAGG scatterer is performed using a separate group of events in which the original gamma quanta are scattered back from the adjacent plastic scatterer and absorbed by the GAGG scintillator. According to the kinematics of such  Compton scattering, the energy deposition in GAGG corresponds to $170.5$ keV. As shown in the insert in figure \ref{fig:6}, right, the extended energy spectrum in GAGG contains a prominent peak, which is used for energy calibration of GAGG scatterer every 2 hours of data acquisition.  Obviously, these calibration events with the absorption of scattered photons in GAGG have no hits in the NaI(Tl) counters of the corresponding arm of the setup and cannot be used in physical analysis. 

\section{Performance of polarimeters}

\begin{figure}[htbp]
\centering 
\includegraphics[width=.47\textwidth]{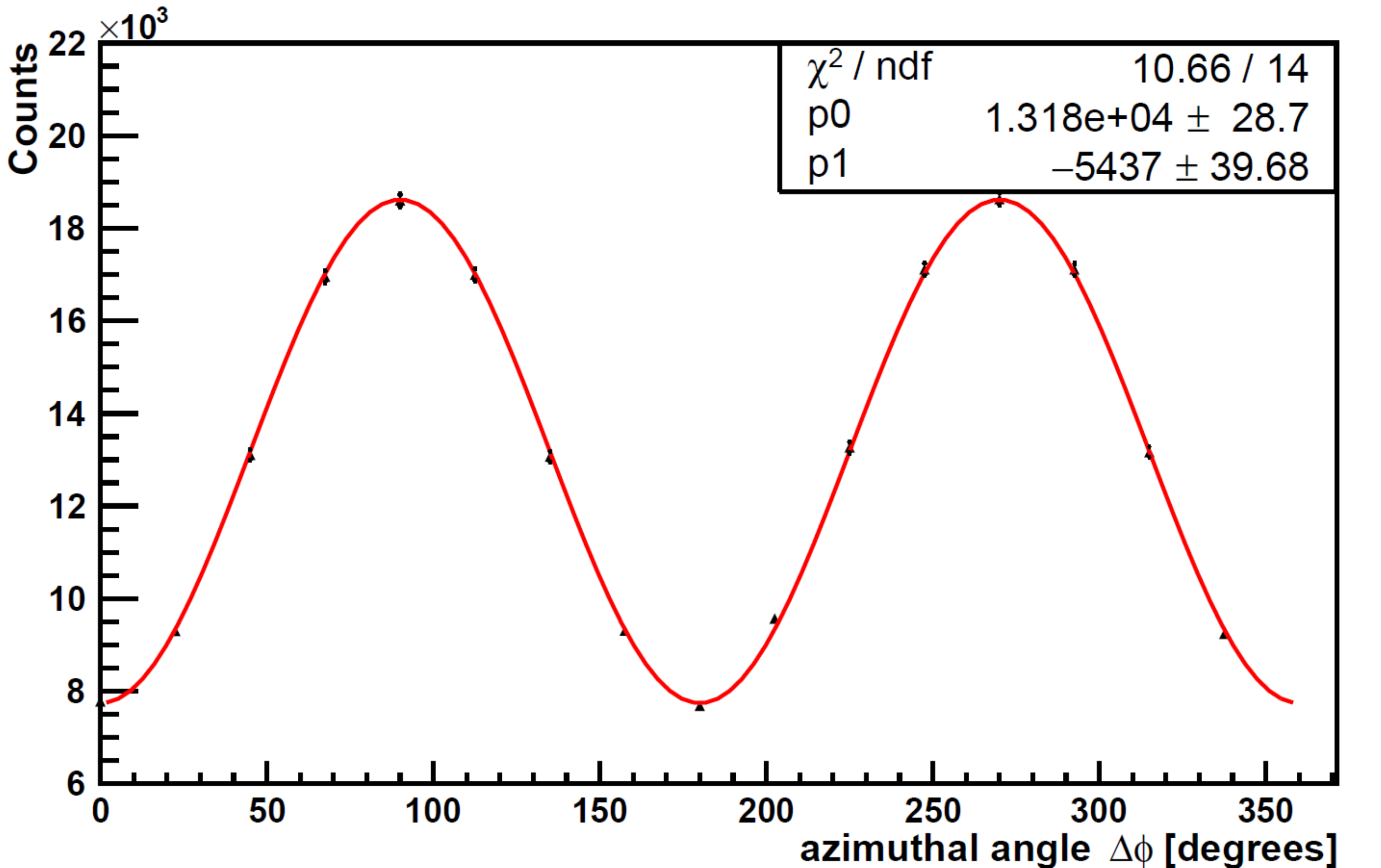}
\qquad
\includegraphics[width=.47\textwidth]{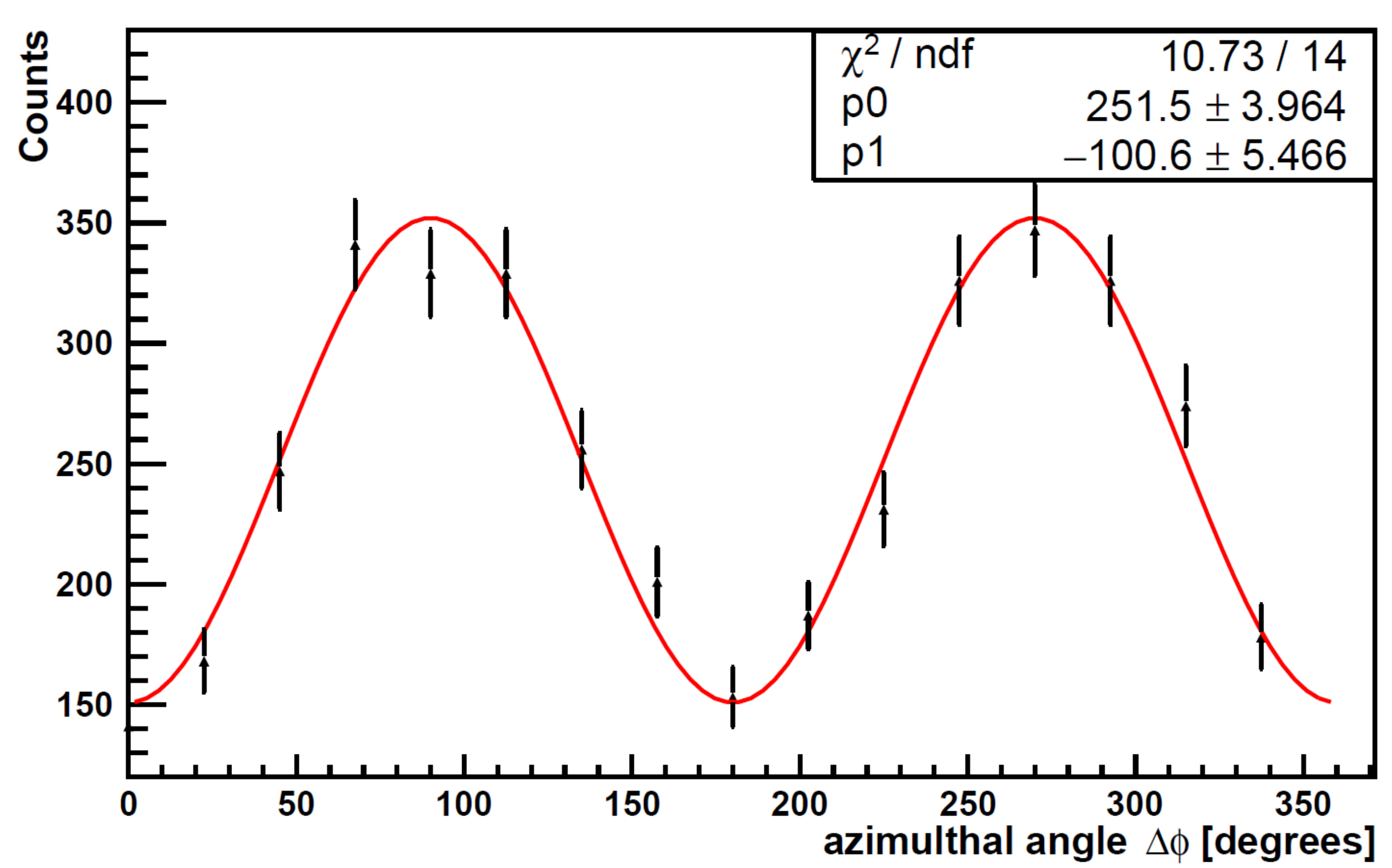}
\caption{ Left - the dependence of the number of photons registered in the NaI(Tl) counters on the azimuthal angle between these photons. The events in entangled state are selected. Right - the same dependence for the decoherent events with energy deposition in GAGG between 10 keV and 40 keV. The solid lines correspond to the fit function $p_0+p_1\cdot\cos(2\Delta \phi)$.}
\label{fig:7}
\end{figure}

During two months of the setup operation, about $2\times10^5$ events with the annihilation photons in entangled states were selected. The dependence of the number of scattered photons registered in the NaI(Tl) counters on the azimuthal angle between these photons is shown in the  figure \ref{fig:7}, left panel.   Since this azimuthal angle equals to $\Delta \phi$ in Eq. \ref{eq:prob}, counts can be approximated by the function $f(\theta)=A-B\cdot\cos(2\Delta \phi)$. 
Normalized value of $\chi^2/\nu= 0.8$ confirms the agreement between the experimental measurements and the theoretical predictions. 
Sensitivity of the setup to the polarization measurements is characterized by the polarization modulation factor $\mu$, see Eq. \ref{eq:modul}.  By definition, the modulation factor is $\mu=B/A$ with a measured value of $0.41$, which is about 15\% less than the maximum theoretical value of 0.48.

For the entangled photons with mutually orthogonal polarizations modulation factor is a product of analyzing powers of the corresponding Compton polarimeters.
Since all polarimeters in the setup are equivalent, the analyzing power of each polarimeter equals to 0.64 and only 7\% less than maximum theoretical value of 0.69.

 The smaller modulation factor is explained by the finite geometry of NaI(Tl) counters and plastic scatterers that leads to the gamma detection in a wide range of scattering angles from $80^\circ$ to $100^\circ$, while the optimum angle equals to $82^\circ$. At the same time, such a large range of detected scattering angles ensures the statistical accuracy of the measurements, which is especially important for decoherent photons. Note, that the possible contribution of events in decoherent quantum states does not affect the value of the modulation factor. This is illustrated in figure \ref{fig:7}, right, which shows the angular dependence of the number of detected pairs of photons in decoherent states. The events with energy deposition in GAGG intermediate scatterer from 10 keV to 40 keV are selected. These events with small scattering angles have minimum impact on the momenta of initial annihilation photons. As one can see, the modulation factors for these and entangled events are practically the same. This result is consistent with the calculations in theoretical paper \cite{hiesmayr} where identical kinematics of Compton scattering for the annihilation photons in entangled and decoherent states are predicted.

\section{Summary}

The experimental setup for studying the Compton scattering of annihilation photons in entangled and decoherent quantum states is in operation now at the Institute for Nuclear Research of RAS, Moscow. Two sets of Compton polarimeters have large solid angle coverage which is necessary for the acquisition of statistically significant number of decoherent events.  These events are identified by pre-scattering in the GAGG scintillator prior to polarimeter measurement. The rotational symmetry of the setup makes it possible to compensate for systematic errors caused by the possible inefficiency of the detectors and positioning inaccuracies. The modulation factor and the analyzing power of Compton polarimeters were estimated from the angular distributions of scattered photons. It turns out that these distributions are the same for both entangled and decoherent quantum states of annihilation photons. This indicates the identical kinematics of Compton scattering regardless of the type of quantum state.


\begin{thebibliography}{99}
\bibitem{bohm}
D.Bohm and Y. Aharonov, 
\emph{ Discussion of experimental proof of the paradox of
Einstein, Rosen, and Podolsky},
Phys. Rev., { \bf 108} (1957) 1070.

\bibitem{pryce}
Pryce, M., Ward, J., 
\emph{Angular Correlation Effects with Annihilation Radiation}, 
Nature, {\bf 160} (1947) 435.

\bibitem{snyder}
H. Snyder et al.,
\emph{Angular correlation of scattered annihilation radiation}, 
Phys. Rev., {\bf 73} (1948) 440.

\bibitem{kasday}
L. R. Kasday, J. Ullman, and C. S. Wu, 
\emph{Angular correlation of compton-scattered annihilation photons and hidden variables},
Nuov Cim B, { \bf 25} (1975) 633.

\bibitem{bruno}
Bruno, M., D'Agostino, M. and Maroni,
\emph{Measurement of linear polarization of positron annihilation photons}, 
Nuov Cim B, {\bf 40} (1977) 143.

\bibitem{langhoff}
Langhoff, H.,
\emph{Die Linearpolarisation der Vernichtungsstrahlung von Positronen}, 
Z. Physik, {\bf 160} (1960) 186.

\bibitem{caradonna}
Caradonna, P., Reutens, D., Takahashi, T., Takeda, S. and Vegh, V.,
\emph {Probing entanglement in Compton interactions}, J. Phys. Commun., 
{\bf 3} (2019) 105005.

\bibitem{kozuljevich}
Kožuljević AM, Bosnar D, Kuncic Z, Makek M, Parashari S, Žugec P., 
\emph {Study of Multi-Pixel Scintillator Detector Configurations for Measuring Polarized Gamma Radiation}, Condensed Matter, {\bf 6} (2021) 43. 

\bibitem{toghyani}
Toghyani, M., Gillam, J., McNamara, A. and Kuncic, Z.,
\emph{Polarisation-based coincidence event discrimination: An in silico study towards a feasible scheme for Compton-PET},
Phys. Med. Biol., {\bf 61} (2016) 5803.
 
\bibitem{watts}
Watts, D.P., Bordes, J., Brown, J.R. et al.,
\emph{Photon quantum entanglement in the MeV regime and its application in PET imaging},
Nat. Commun., { \bf 12} (2021) 2646.

\bibitem
{pet1} Moskal, P., Krawczyk, N., Hiesmayr, B.C. et al.,
\emph{ Feasibility studies of the polarization of photons beyond the optical wavelength regime with the J-PET detector},
 Eur. Phys. J. {\bf C 78} (2018) 970, 1809.10397.
 
 \bibitem
{pet2} Moskal, P., 
\emph{ Towards total-body modular PET for positronium and quantum entanglement imaging},
 2018 IEEE Nuclear Science Symposium and Medical Imaging Conference Proceedings (NSS/MIC) (2018), pp. 1-4, doi: 10.1109/NSSMIC.2018.8824622.

\bibitem{hiesmayr}
Hiesmayr, B.C., Moskal, P.,
\emph {Witnessing Entanglement In Compton Scattering Processes Via Mutually Unbiased Bases},Sci. Rep., 
{\bf 9} (2019) 8166. 


\bibitem{nishina}
Klein, O., Nishina, Y.,
\emph {Über die Streuung von Strahlung durch freie Elektronen nach der neuen relativistischen Quantendynamik von Dirac}, 
Z. Physik {\bf 52} (1929) 853.


\bibitem{knights}
P Knights, F Ryburn, G Tungate, K Nikolopoulos, 
\emph {Studying the effect of polarisation in Compton scattering in the undergraduate laboratory}, Eur. J. Phys. {\bf 39} (2018) 025203, 1711.06510.

\bibitem{source}
B.L.Zhuikov, V.M.Kokhanyuk, N.A.Konyakhin, J.Vincent, 
\emph {Target irradiation facility and targetry development at 160 MeV proton beam of Moscow linac}, Nucl. Instr. Meth.
{\bf A438} (1999) 173.


\bibitem{garwin}
R. L. Garwin, 
\emph {Thermalization of positrons in metals}, Phys. Rev. {\bf 91}  (1953) 1571.


\bibitem{positron}
P. Hautojarvi and A. Vehanen,
\emph { in Positrons in Solids, edited by P. Hautojarvi, Springer-Verlag, Berlin} (1979) 14.

\bibitem{afi}
\emph {AFI electronics webpage}, http://afi.jinr.ru.

\bibitem{JPET} Kamińska, D., Gajos, A., Czerwiński, E. et al., 
\emph {A feasibility study of ortho-positronium decays measurement with the J-PET scanner based on plastic scintillators}, Eur. Phys. J. {\bf C76} (2016) 445, 1607.08588.



\FloatBarrier

\end{thebibliography}
\end{document}